\documentclass[letter]{aa} 
\usepackage{graphicx}
\usepackage{natbib}
\bibpunct{(}{)}{;}{a}{}{,}
\usepackage{txfonts}
\newcounter{Rco}
\newcommand{\Ionst}[1]{\setcounter{Rco}{#1}\Roman{Rco}}
\newcommand{\Ion}[2]{\mbox{#1\ {\scriptsize\Ionst{#2}}}}
\newcommand{\Ionw}[3]{\mbox{#1\ {\scriptsize\Ionst{#2}}~$\lambda\,#3$\,\AA}}

\newcommand{\logg}{\mbox{$\log g$}}
\newcommand{\loggw}[1]{\mbox{$\log g\hspace{-0.5mm} =\hspace{-0.5mm}  #1$}}
\newcommand{\mc}[3]{\multicolumn{#1}{#2}{#3}}
\newcommand{\kK}{\mbox{\rm kK}}

\newcommand{\sA}[1]{\mbox{(Fig.\,\ref{#1})}}

\newcommand{\spm}{\mbox{\raisebox{0.20em}{{\tiny \hspace{0.2mm}\mbox{$\pm$}\hspace{0.2mm}}}}}

\newcommand{\Teff}{\mbox{$T_\mathrm{eff}$}}
\newcommand{\Teffp}{\mbox{$T_\mathrm{eff}^\mathrm{pri}$}}
\newcommand{\Teffw}[1]{\mbox{$\Teff\hspace{-0.5mm}=\hspace{-0.5mm}#1\,\mathrm{K}$}}
\newcommand{\hs}[1]{\hbox{}\hspace{#1}}
\newcommand{\emptytwo}{\multicolumn{2}{c}{~}}
\newcommand{\aador}{\object{AA\,Dor}}
\newcommand{\aadorf}{\object{AA\,Doradus}}
\newcommand{\lb}{\object{LB\,3459}}
\def\tmap{\emph{TMAP}}

\begin{document}
   \title{On the sdOB primary of the post common-envelope binary \aadorf\ (\lb)\thanks
          {Based on observations made with ESO Telescopes at the Paranal Observatories under
           programme ID 66.D-1800.
          }
         }

   \titlerunning{Non-LTE spectral analysis of the sdOB primary of \aador}

   \author{S. Klepp
           \and
           T. Rauch}

   \institute{Institute for Astronomy and Astrophysics,
              Kepler Center for Astro and Particle Physics,
              Eberhard Karls University,
              Sand 1,
              D-72076 T\"ubingen,
              Germany,
              \email{rauch@astro.uni-tuebingen.de}
             }

   \date{Received March 14, 2011; accepted June 6, 2011}

 
  \abstract
   {\aador\ is an eclipsing, post common-envelope binary with an sdOB-type primary  
    and a low-mass secondary. 
    Eleven years ago, an NLTE spectral analysis showed a discrepancy in the surface gravity 
    that was derived by radial-velocity and light-curve analysis, \loggw{5.21\pm 0.1} ($\mathrm{cm/sec^2}$) and
    \loggw{5.53\pm 0.03}, respectively.
   }
   {We aim to determine both the effective temperature and surface gravity of \aador\ precisely
    from high-resolution, high-S/N observations taken during the occultation of the secondary.
   }
   {We calculated an extended grid of metal-line blanketed, state-of-the-art, non-LTE model atmospheres
    in the parameter range of the primary of \aador. Synthetic spectra calculated from
    this grid were compared to optical observations.
   }
   {We verify \Teffw{42000\pm 1000} from our former analyses and determine 
    a higher \loggw{5.46\pm 0.05}. The main reason are new Stark-broadening tables
    that were used for calculating of the theoretical Balmer-line profiles.
   }
   {Our result for the surface gravity agrees 
    with the value from light-curve analysis within the error limits, 
    thereby solving the so-called gravity problem in \aador.
   }


   \maketitle

\section{Introduction}
\label{sect:intro}

\aador\ is a close, eclipsing, post common-envelope binary system with an sdOB-type primary star 
and an unseen low-mass companion. The orbital period 
is 
0.261\,539\,7363 (4)\,d \citep{kilkenny2011} 
and the inclination is 
$i=89\fdg 21\pm 0\fdg 30$ \citep{Hilditch03}. 
A detailed introduction to the system and previous analyses is given in \citet{rauch2004}, 
and we summarize results of previous spectral analyses of the primary in Table\,\ref{tab:previous}.

\begin{table*}[ht!]
\caption{Effective temperature and surface gravity of the primary of \aador,
         determined in previous and the present spectral analyses.}
\label{tab:previous}
\begin{tabular}{r@{}lr@{.}lr@{.}lr@{.}lr@{.}lr@{.}lrl}
\hline
\hline
\noalign{\smallskip}
\multicolumn{2}{c}{\Teff} & 
\multicolumn{2}{c}{\logg} & 
\multicolumn{2}{c}{He} & 
\multicolumn{2}{c}{$v_\mathrm{rot}^\mathrm{pri}$} & 
\multicolumn{2}{c}{$M_\mathrm{pri}$} & 
\multicolumn{2}{c}{$M_\mathrm{sec}$)} & 
method & 
reference \\
\multicolumn{2}{c}{(K)} & 
\multicolumn{2}{c}{(cm/sec$^2$)} & 
\multicolumn{2}{c}{(mass fraction)} & 
\multicolumn{2}{c}{(km/sec)} & 
\multicolumn{2}{c}{($M_\odot$)} & 
\multicolumn{2}{c}{($M_\odot$)} & 
& 
\\
\noalign{\smallskip}
\hline
\noalign{\smallskip}
41000&                   & \hs{7mm}5&4                    & \hs{8mm}0&28   & \emptytwo    & \emptytwo    & \emptytwo     &  LTE$^1$    & \citet{kudritzki1976} \\
44200&                   &         5&2                    &         0&28   & \emptytwo    & \emptytwo    & \emptytwo     & NLTE$^1$    & \citet{kudritzki1976} \\
41700&                   &         5&9                    &         0&8    & \emptytwo    & \emptytwo    & \emptytwo     &  LTE$^1$    & \citet{kudritzki1976} \\
42000&                   &         5&7                    &         0&8    & \emptytwo    & \emptytwo    & \emptytwo     & NLTE$^1$    & \citet{kudritzki1976} \\
40000&$^{+3000}_{-2000}$ &         5&3\hs{1.5mm}$\pm 0.2$ &         0&012  & \emptytwo    &         0&3  &         0&04  & NLTE$^1$    & \citet{kudritzkietal1982} \\
42000&$\pm$1000          &         5&21$\pm 0.1$          &         0&0032 & \hs{5mm}34&0 & \hs{5mm}0&33 & \hs{5mm}0&066 & NLTE$^2$    & \citet{rauch2000} \\
42000&$\pm$1000          &         5&30$\pm 0.1$          &         0&0032 &         35&0 & \emptytwo    & \emptytwo     & NLTE$^3$    & \citet{fleigetal2008} \\
37800&$\pm$\hs{1.5mm}500 &         5&51$\pm 0.05$         &         0&005  &         30&0 &         0&51 &         0&085 &  LTE$^4$    & \citet{muelleretal2010} \\
42000&$\pm$1000          &         5&46$\pm 0.05$         &         0&0027 &         30&0 &         0&47 &         0&079 & NLTE$^5$    & this work \\
\noalign{\smallskip}
\hline
\noalign{\smallskip}
\end{tabular}\\
{\tiny
$^1$ H+He models, two grids with fixed $N_\mathrm{He}$/$N_\mathrm{H}$ ratios only, $1^\mathrm{st}$ investigation of NLTE effects, no errors given \\
$^2$ H+He+C+N+O+Mg+Si+Fe+Ni, Fe+Ni data from \citet{kurucz1991}, optical spectra, assumed bound rotation ($v_\mathrm{rot}^\mathrm{pri} =45.7\,\mathrm{km/sec}$) \\
$^3$ H+He+C+N+O+Mg+Si+P+S+Ca+Sc+Ti+V+Cr+Mn+Fe+Co+Ni, Ca-Ni data from \citet{kurucz1991}, optical and FUV spectra \\
$^4$ H+He metal enhanced ($z$ = ten times solar), optical spectra \\
$^5$ H+He+C+N+O+Mg+Si+P+S+Ca+Sc+Ti+V+Cr+Mn+Fe+Co+Ni, Ca-Ni data from \citet{kurucz2009}, optical spectra, $v_\mathrm{rot}^\mathrm{pri}$ from \citet{muelleretal2010} \\
}
\end{table*}

\citet{rauch2000} encountered the problem of his spectroscopically determined surface gravity \loggw{5.21\pm 0.1}
not matching \loggw{5.53\pm 0.03} determined from light-curve analysis
\citep{Hilditch96}. 
\citet{Hilditch03} present an improved photometric model and derive \loggw{5.45 - 5.51}.
The reason for the \logg\ discrepancy is unknown. \citet{fleigetal2008} find
a slightly higher \loggw{5.3\pm 0.1} but the discrepancy remains.
A recent analysis by \citet{muelleretal2010} has apparently solved the \logg\ problem by
finding \loggw{5.51\pm 0.05}.

\begin{figure}
\centering
\includegraphics[width=\columnwidth]{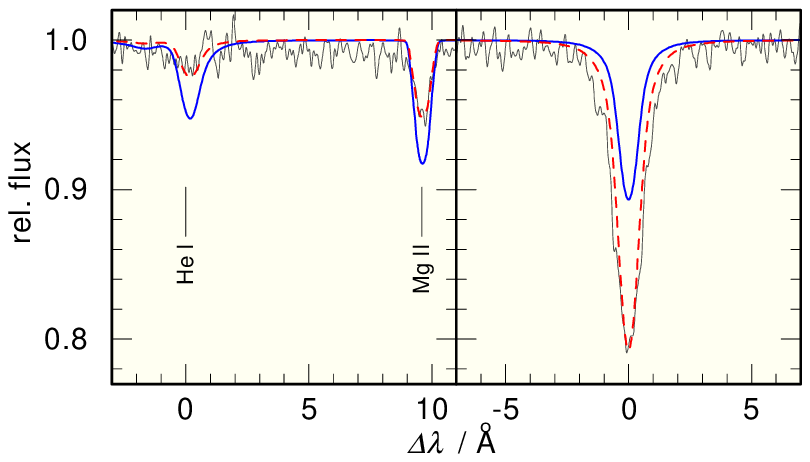}
\caption{Comparison of our synthetic spectra 
         (full, blue line: \Teffw{37800}, \loggw{5.51};
              dashed, red: \Teffw{42000}, \loggw{5.46}; He = 0.0027 by mass)
          around 
          \Ionw{He}{1}{4471} and \Ionw{Mg}{2}{4481} (left) and
          \Ionw{He}{2}{4686} (right)
          with the observation. The models are convolved with a rotational
          profile corresponding to $v_\mathrm{rot} = 30\,\mathrm{km/sec}$.
          Models and observation are smoothed with
          a Gaussian (0.1\,\AA\ FWHM) for clarity.
           }
\label{fig:hei}
\end{figure}

\citet{muelleretal2010} do not consider that the \Ion{He}{1} lines 
(as well as other lines of low-ionized species, e.g\@. of \Ion{Mg}{2}, Fig\@.\,\ref{fig:hei}) 
are too strong in the models
at their favored para\-meters \Teffw{37800\pm 500} and \loggw{5.51\pm 0.05}, as demonstrated in Fig.\ref{fig:hei}.
Increasing the He abundance to better fit \Ionw{He}{2}{4686} also results in a much stronger
\Ionw{He}{1}{4471}, which then disagrees with the observation.
We mention that \citet{fleigetal2008} evaluate the ionization equilibria of
\Ion{C}{3} / \Ion{C}{4}, 
\Ion{N}{3} / \Ion{N}{4}, 
\Ion{O}{3} / \Ion{O}{4},
\Ion{P}{4} / \Ion{P}{5}, and
\Ion{S}{4} / \Ion{S}{5}
in the FUV wavelength range and find \Teffw{42000\pm 1000}
to agree with the higher \Teff\ concluded from the \Ion{He}{1} lines.

Since \citet[][http://kurucz.harvard.edu/atoms.html]{kurucz2009} has substantially
extended his database, and the model atoms in our 
T\"ubingen Model-Atom Database 
(\emph{TMAD}\footnote{http://astro.uni-tuebingen.de/\raisebox{0.3em}{{\tiny $\sim$}}TMAD/TMAD.html}) 
have been updated as well, we decided to calculate an improved, extended, state-of-the-art
NLTE model-atmosphere grid. This grid is described in Sect.\,\ref{sect:NLTE}.
The re-analysis of our UVES spectra 
\citep[105 \'a 180 \,sec, 
which in total cover one orbital period and which were also used by][]{muelleretal2010}
that were obtained in 2001 at the VLT is described in Sect.\,\ref{sect:results}.
We conclude in Sect.\,\ref{sect:conclusions}.

\section{Atomic data and model-atmosphere grid}
\label{sect:NLTE}

The model atmospheres used here were calculated with the 
T\"ubingen Model-Atmosphere Package \citep[\tmap]{werneretal2003}.
The models are plane-parallel, in hydrostatic and radiative equilibrium.
\tmap\ uses the occupation-probability formalism of
\citet{hm1988} that was generalized to NLTE conditions by
\citet{hea1994}.
\tmap\  considers opacities of 
H+He+C+ N+O+Mg+Si+P+S using classical model atoms, and Ca+Sc+Ti+V+Cr+Mn+Fe+Co+Ni uses
a statistical approach \citep{rd2003}.
All model atoms used in our calculations 
were updated to the most recent atomic data (Sect.~\ref{sect:intro}),
and 530 levels are treated in NLTE with
771 individual lines (from H\,-\,S) and 
19\,957\,605 lines of Ca\,-\,Ni from Kurucz' line lists \citep{kurucz2009} combined to 636 superlines.
The element abundances are summarized in Table\,\ref{tab:abund}.

\onltab{2}{
\begin{table}
\caption{Element abundances in our model-atmosphere grid.}
\label{tab:abund}
\begin{tabular}{cr@{.}lr@{.}lr@{.}lr@{.}l}
\hline
\hline
\noalign{\smallskip}
element &
\multicolumn{2}{c}{mass} &
\multicolumn{2}{c}{number} &
\multicolumn{2}{c}{~} &
\multicolumn{2}{c}{~} \\
 X &
\multicolumn{2}{c}{fraction} &   
\multicolumn{2}{c}{fraction} &    
\multicolumn{2}{c}{[$\epsilon_\mathrm{X}$]$^\ast$} &    
\multicolumn{2}{c}{[X]$^{\ast\ast}$} \\
\noalign{\smallskip}
\hline
\noalign{\smallskip}
  H     & 9&939E$-$1 & 9&992E$-$1 &  12&144 & $+$0&130 \\
  He    & 2&686E$-$3 & 6&801E$-$4 &   8&977 & $-$1&967 \\
  C     & 1&777E$-$5 & 1&499E$-$6 &   6&320 & $-$2&124 \\
  N     & 4&145E$-$5 & 2&998E$-$6 &   6&621 & $-$1&223 \\
  O     & 1&009E$-$3 & 6&393E$-$5 &   7&950 & $-$0&754 \\
  Mg    & 4&078E$-$4 & 1&700E$-$5 &   7&375 & $-$0&240 \\
  Si    & 3&049E$-$4 & 1&100E$-$5 &   7&186 & $-$0&339 \\
  P     & 5&197E$-$6 & 1&700E$-$7 &   5&375 & $-$0&050 \\
  S     & 3&241E$-$6 & 1&024E$-$7 &   5&155 & $-$1&980 \\
  Ca    & 5&994E$-$5 & 1&515E$-$6 &   6&325 & $-$0&030 \\
  Sc    & 3&694E$-$8 & 8&327E$-$0 &   3&065 & $-$0&100 \\
  Ti    & 2&784E$-$6 & 5&894E$-$8 &   4&915 & $-$0&050 \\
  V     & 3&731E$-$7 & 7&421E$-$9 &   4&015 & $+$0&070 \\
  Cr    & 1&663E$-$5 & 3&240E$-$7 &   5&655 & $+$0&001 \\
  Mn    & 9&877E$-$6 & 1&828E$-$7 &   5&405 & $-$0&040 \\
  Fe    & 1&153E$-$3 & 2&091E$-$5 &   7&465 & $-$0&050 \\
  Co    & 3&591E$-$6 & 6&174E$-$8 &   4&935 & $-$0&069 \\
  Ni    & 3&482E$-$4 & 6&009E$-$6 &   6&923 & $+$0&689 \\
\noalign{\smallskip}
\hline
\end{tabular}\\
{\tiny
\hbox{}\hspace{1mm}$^\ast$:       $\log \left(\epsilon_\mathrm{i} / \epsilon_\odot\right)$, 
               $\log \sum_\mathrm{i} \mu_\mathrm{i} \epsilon_\mathrm{i} = 12.15$,
               \citep[cf\@.][]{holweger1979}\\
$^{\ast\ast}$: log[abundance/solar abundance] \citep[solar values from][]{asplundetal2009}
}
\end{table}
}

\begin{figure}
\centering
\includegraphics[width=\columnwidth]{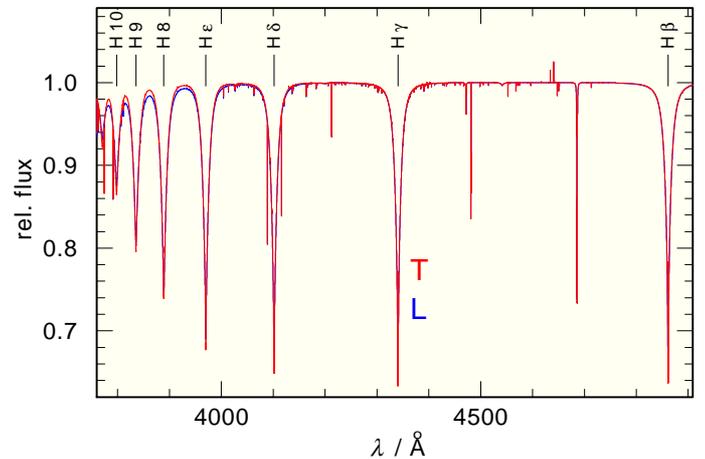}
\caption{Synthetic spectrum calculated from a 
         \Teffw{42000} and \loggw{5.45} model with
         different Stark line-broadening tables 
         (L, blue line: \citet{lemke1997},
          T, red: \citet{tremblaybergeron2009}, see text).
        }
\label{fig:stark}
\end{figure}

The model-atmosphere grid spans 
\Teffw{35000 - 49000} ($\Delta$\,\Teffw{500}) and 
\loggw{5.15 - 6.20} ($\Delta$\,\loggw{0.05}). In total this makes 638 models.
Spectral energy distributions (SEDs) were calculated
using the most recent line broadening data, e.g\@.
H\,{\sc i} line-broadening has changed in \emph{TMAP} since \citet{fleigetal2008}
presented their analysis of \aador. The reason is that \citet{repolust:2005} found 
an error in the H\,{\sc i} line-broadening tables (for high members of the spectral 
series only) by \citet{lemke1997} that were used before. These were substituted 
by a Holtsmark approximation. In addition, \citet{tremblaybergeron2009} provide new, 
parameter-free Stark line-broadening tables for H\,{\sc i} considering non-ideal effects. 
These replaced Lemke's data for the lowest ten members of the H\,{\sc i} Lyman and Balmer series.
In the parameter range of \aador, the new broadening tables have a significant impact 
on the line wings of higher Balmer-series members (narrower for H\,$\epsilon$ and higher,
Fig.\,\ref{fig:stark}). As a consequence, our analysis results in a higher
\logg\ (Sect.\,\ref{sect:results}).

In the framework of the Virtual
Observatory\footnote{http://www.ivoa.net} (\emph{VO}), all these SEDs ($\lambda - F_\lambda$)  
are available in \emph{VO} compliant form
via the \emph{VO} service 
\emph{TheoSSA}\footnote{http://vo.ari.uni-heidelberg.de/ssatr-0.01/TrSpectra.jsp?}
provided by the \emph{German Astrophysical Virtual Observatory}
(\emph{GAVO}\footnote{http://www.g-vo.org}).

\section{Analysis and results}
\label{sect:results}

\begin{figure}
\centering
\includegraphics[width=\columnwidth]{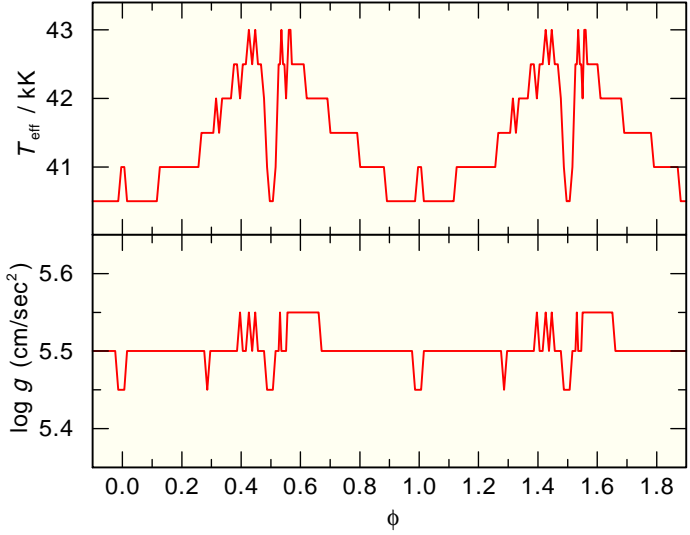}
\caption{Phase-dependent, best-fitting grid model determined by a $\chi^2$ fit.
         $\phi = 0.0$ is the transit, $\phi = 0.5$ is the occultation of the secondary.
         (The steps in \Teff\ and \logg\ represent the grid spacing.)
        }
\label{fig:phase}
\end{figure}

The light curve of \aador\ exhibits a reflection effect \citep[e.g\@.][]{Hilditch96}
that amounts to about 0.06\,mag in the optical. To analyze the pure primary
spectrum, we selected only those four observations that were taken closest to the occultation
of the secondary. These were co-added in order to improve the S/N.
In Fig.\,\ref{fig:phase}, we show a $\chi^2$ fit to all single UVES spectra. 
Our $\chi^2$ fit excludes the inner line cores of H\,$\beta$ and H\,$\gamma$, as
well as obviously bad data points \citep[quality flags given by][]{muelleretal2010}.
Both the occultation (at $\phi = 0.5$) 
and
the transit ($\phi = 0.0$) 
of the secondary
are clearly visible in the determination of
\Teff\ and \logg. 
Compared to a similar $\chi^2$ fit of \citet[][their Fig.\,3]{muelleretal2010}, we find 
the same \loggw{5.45} but a
significantly higher \Teffw{40500} than for \Teffw{37800}.

\begin{figure}
\centering
\includegraphics[width=\columnwidth]{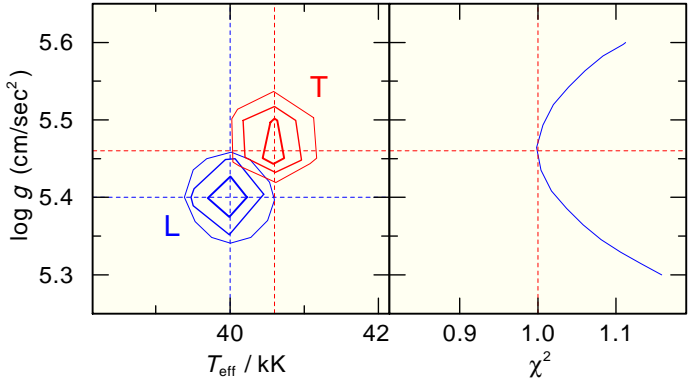}
\caption{Left:  Formal 1\,$\sigma$, 2\,$\sigma$, and 3\,$\sigma$
                contour lines of our $\chi^2$ fits in the \Teff\ - \logg\ plane.
         Right: Reduced {\boldmath $\chi^2$\unboldmath} of our \Teffw{42000} models depending on \logg.
}
\label{fig:contour}
\end{figure}

For the analysis, 
we perform a detailed comparison in the classical way ($\chi$-by-eye) and, 
for comparison in analogy to \citet{muelleretal2010} with a
$\chi^2$ fit, used the same wavelength limits (Table\,\ref{tab:limits}) and lines, 
H\,$\beta$ - H\,11 and \Ionw{He}{2}{4686}. 
Our $\chi^2$ fit yields \Teffw{40600\pm 100} and \loggw{5.46^{+0.04}_{-0.02}} 
(T in Fig.\,\ref{fig:contour}).
These errors are formal $1\,\sigma$ errors, and $\sigma$ was calculated from the 
deviation of the $\chi^2_\mathrm{min}$ model from the observed spectrum used in
the $\chi^2$ fit. 
Compared to a similar $\chi^2$ fit with SEDs that were calculated with
the previously used Stark broadening tables of \citet[][L in Fig.\,\ref{fig:contour}]{lemke1997}, 
there is a significant deviation of
$\Delta T_\mathrm{eff} = 600\,\mathrm{K}$
and
$\Delta \log g = 0.06$.

\onltab{3}{
\begin{table}
\caption{Lines and wavelength intervals used for our $\chi^2$ fits.}
\label{tab:limits}
\begin{tabular}{cr@{$,$}lrr@{$,$}l}
\hline
\hline
\noalign{\smallskip}
line & \multicolumn{2}{c}{$\Delta\lambda$ } & line & \multicolumn{2}{c}{$\Delta\lambda$} \\
     & \multicolumn{2}{c}{(\AA)}            &      & \multicolumn{2}{c}{(\AA)}             \\
\hline
\noalign{\smallskip}                     
H\,$\beta$         & [$-$50 & +50]          & \Ionw{He}{1}{4471} & [\hs{1.6mm}$-$5  & \hs{1.6mm}+5] \\
H\,$\gamma$        & [$-$40 & +40]          & \Ionw{He}{2}{4686} & [\hs{1.6mm}$-$5  & \hs{1.6mm}+5] \\
H\,$\delta$        & [$-$30 & +30]          \\
H\,$\epsilon$      & [$-$20 & +20]          \\
H\,8               & [$-$20 & +20]          \\
H\,9               & [$-$10 & +10]          \\
H\,10              & [$-$10 & \hs{1.6mm}+7] \\
H\,11              & [$-$10 & \hs{1.6mm}+8] \\
\hline
\end{tabular}
\end{table}
}

\begin{figure}
\centering
\includegraphics[width=\columnwidth]{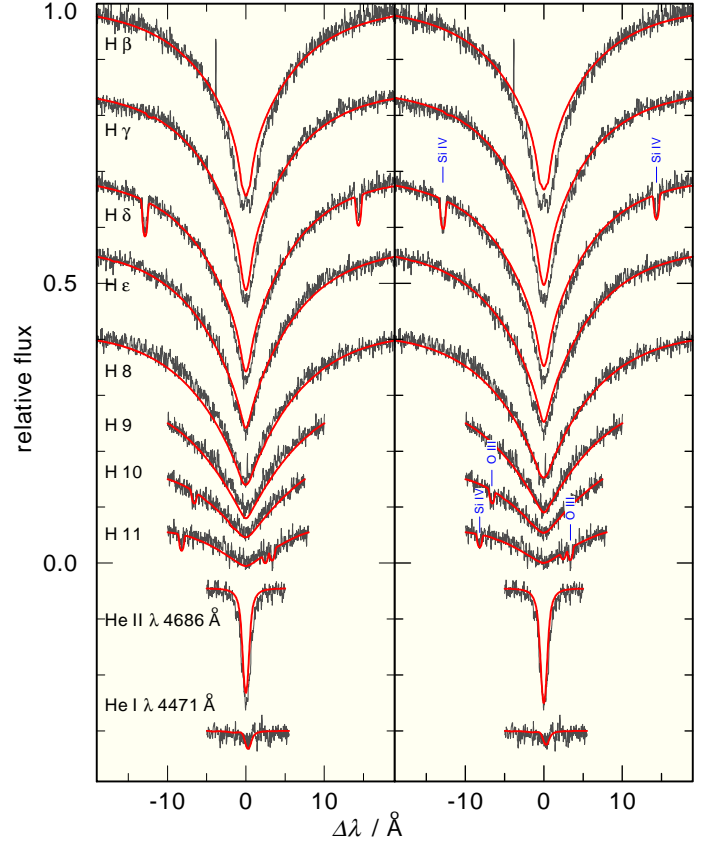}
\caption{Comparison of synthetic line profiles of H and He lines calculated from a 
         \Teffw{40600} and \loggw{5.46} model (left) and a
         \Teffw{42000} and \loggw{5.46} model (right) with
          the observation.}
\label{fig:final}
\end{figure}

A comparison of the best-fitting model from our $\chi^2$ fit and the
best-fitting \,$\chi$-by-eye with the observations is shown in Fig.\,\ref{fig:final}.
It is obvious that the ionization equilibrium of \Ion{He}{1}\,/\,\Ion{He}{2}
is reproduced not at 
\Teffw{40600} 
but at \Teffw{42000}. 
The theoretical line profiles of 
lower members of the Balmer series (H\,$\beta$ - H\,$\delta$)
do not reproduce the observation perfectly. 
They fit slightly better at \Teffw{40600}.
Thus, a small Balmer-line problem
\citep{nr1994, werner1996}
due to additional metal opacities that are still not considered is apparently present.
The inclusion of \Ionw{He}{1}{4471} (Table\,\ref{tab:limits}) in the $\chi^2$-fit procedure
results in higher 
\Teffw{40700\pm 300}. 
We finally adopt \Teffw{42000\pm 1000} \citep[cf\@.][]{fleigetal2008} and 
\loggw{5.46\pm 0.05}
because the previously evaluated ionization equilibria \citep{rauch2000,fleigetal2008}
are an additional, crucial constraint. 
A $\chi^2$ fit at fixed \Teffw{42000} 
(additional models were calculated with \loggw{5.30 - 5.60} and $\Delta$\,\loggw{0.01})
also has its minimum at \loggw{5.46} (Fig.\,\ref{fig:contour}).

\begin{figure}
\centering
\includegraphics[width=\columnwidth]{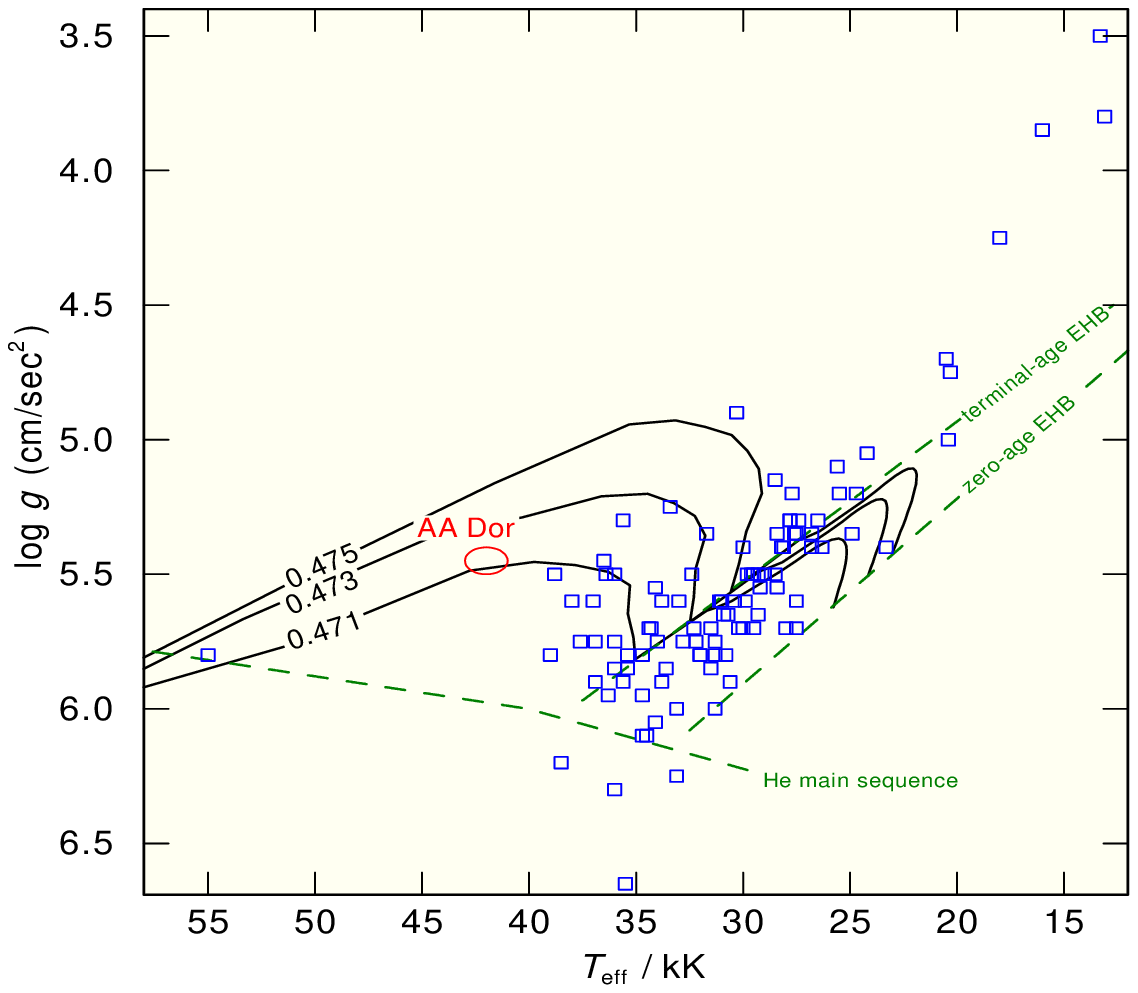}
\caption{Location of \aador\ in the \Teff$-$\logg\
         plane compared to sdBs and sdOBs from \citet{edelmann2003}.
         Post-EHB tracks from \citet[][labeled with the
         respective stellar masses in $M_\odot$]{dormanetal1998} 
         are also shown.
        }
\label{fig:evol}
\end{figure}

A mass of $M_\mathrm{pri} = 0.4714\pm 0.0050\,M_\odot$ is determined by comparing
of \Teff\ and \logg\ with the evolutionary tracks of post-EHB stars (Fig.\,\ref{fig:evol}). From the same evolutionary
calculations, we interpolate the primary's luminosity. 
From our final model, we can determine the spectroscopic distance of \aador\
following \citet{heberetal1984}.
We derive a distance of $d= 352^{+20}_{-23} \,\mathrm{pc}$. 
The parameters of \aador\ are summarized in Tables\,\ref{tab:abund} and \ref{tab:para}.

\begin{table}[ht]
\caption[]{Parameters of \aador\ compared with values of
\citet{Hilditch03}.}
\label{tab:para}
\begin{tabular}{r@{\,/\,}lr@{.}llcr@{.}l@{\,$-$\,}r@{.}l}
\hline
\hline
\noalign{\smallskip}
\multicolumn{5}{c}{this work} & & \multicolumn{4}{c}{\citet{Hilditch03}} \\
\cline{1-5}
\cline{7-10}
\noalign{\smallskip}                 
\Teffp                                & \kK                          &  42&0          & \spm 1                 &~& \multicolumn{4}{c}{~} \\
$\log \mathrm{(}g_\mathrm{pri}$       & $\mathrm{\frac{cm}{s^2})}$   &   5&46         & \spm 0.05              & & \hs{5mm}5&45  & 5&51  \\
\noalign{\smallskip}                                                                                              
$M_\mathrm{pri}$                      & $M_\odot$                    &   0&4714       & \spm 0.0050            & &         0&33  & 0&47  \\
\noalign{\smallskip}                                                                                              
$L_\mathrm{pri}$                      & $L_\odot$                    & \mc{2}{c}{120} & $^{+15}_{-20}$         & & \multicolumn{4}{c}{~} \\
\noalign{\smallskip}                                                                                              
$M_\mathrm{sec}$                      & $M_\odot$                    &   0&0788       & $^{+0.0075}_{-0.0063}$ & &         0&064 & 0&082 \\
\noalign{\smallskip}
$d$                                   & pc                           & \mc{2}{c}{352} & $^{+20}_{-23}$         & & \multicolumn{4}{c}{~} \\
\noalign{\smallskip}
\hline
\end{tabular}
\end{table}

\section{Conclusions}
\label{sect:conclusions}

\begin{figure}
\centering
\includegraphics[width=\columnwidth]{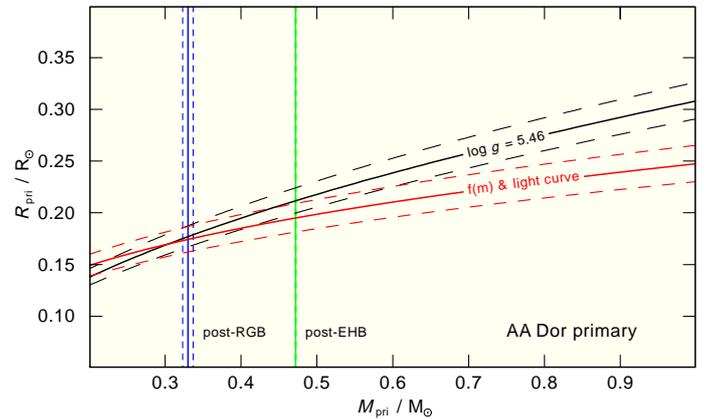}\vspace{-1mm}
\caption{Mass-radius relation for the primary of \aador. 
         The dashed lines show the error ranges.
         The vertical lines show the primary mass, derived from
         comparison with post-RGB \citep{rauch2000}
         and post-EHB (Fig\@.\,\ref{fig:evol})
         evolutionary models.
        }
\label{fig:logg}
\end{figure}

The so-called \logg\ problem in \aador\ is solved (Fig.\,\ref{fig:logg}) and 
our results (Table\,\ref{tab:para}) are in good agreement with the photometric model of \citet{Hilditch03}. 

Four influences were identified on the \logg\ determination.
1) The major impact is
   the improvement in the Stark broadening tables, i.e\@. the difference
   between those of \citet{lemke1997} and of \citet{tremblaybergeron2009}.
   This results in a systematic deviation of 
   $\Delta T_\mathrm{eff} = 600\,\mathrm{K}$
   and
   $\Delta \log g = 0.06$.
2) The reflection effect is now eliminated by using only observed spectra that were
   obtained during the occultation of the secondary (Sect.\,\ref{sect:results}).
3) The improved atomic data makes the model-atmosphere more reliable thanks to a
   fuller consideration of the metal-line blanketing. The temperature stratification
   of the stellar models, however, is only marginally affected.
4) The rotational velocity is lower than previously assumed \citep{muelleretal2010}.
   This only has little influence on the inner line core and is thus important
   for weak and narrow lines like \Ionw{He}{2}{4686} (Fig.\,\ref{fig:hei}).

Since \citet{vuckovicetal2008} identified spectral lines of the secondary in the UVES
spectra and determined a lower limit ($K_\mathrm{sec} > 230\,\mathrm{km/sec}$) of its 
orbital velocity amplitude, both components' masses are known
\citep[$M_\mathrm{pri} = 0.45\,M_\odot$, 
$M_\mathrm{sec} = 0.076\,M_\odot$,][]{vuckovicetal2008}, albeit
with large error bars. 
\citet{muelleretal2010} used the velocity amplitudes of both components 
($K_\mathrm{pri} = 40.15\pm 0.11\,\mathrm{km/sec}$, $K_\mathrm{sec} = 240\pm 20\,\mathrm{km/sec}$) 
to derive the masses
$M_\mathrm{pri} = 0.51^{+0.125}_{-0.108}\,M_\odot$ and
$M_\mathrm{sec} = 0.085^{+0.031}_{-0.023}\,M_\odot$.
This rules out a post-RGB scenario because post-RGB masses are significantly lower.
The solution from mass function f(m) and light curve analysis, however,
intersects with our result of $\log g = 5.46\pm 0.05$ \sA{fig:logg} even for the higher post-EHB mass \sA{fig:evol}.

From our mass determination of $M_\mathrm{pri} = 0.4714\pm 0.0050\,M_\odot$,
we calculated ($M_\mathrm{pri}K_\mathrm{pri} = M_\mathrm{sec}K_\mathrm{sec}$) 
the secondary's mass of $M_\mathrm{sec} = 0.0725 - 0.0863\,M_\odot$.
Since the hydrogen-burning mass limit is about $0.075\,M_\odot$ \citep{chabrierbaraffe1997,chabrieretal2000},
the secondary may either be a brown dwarf or a late M dwarf.

\begin{acknowledgements}
We thank an anonymous referee, the editor R\@. Napiwotzki, and K\@. Werner 
for comments and discussions that helped to improve the paper.
The UVES spectra used in this analysis were obtained as part of an ESO Service Mode run,
proposal 66.D-1800.
This research made use of the SIMBAD Astronomical Database, operated at the CDS, Strasbourg, France.
We thank the \emph{GAVO} team for support.
TR is supported by the German Aerospace Center (DLR) under grant 05\,OR\,0806. 
\end{acknowledgements}


\begin{thebibliography}{}

\bibitem[Asplund et al\@.(2009)Asplund et al\@.]{asplundetal2009}
         Asplund, M\@., Grevesse, N., \& Sauval, A\@. J\@.
         2009,
         \araa, 47, 481

\bibitem[Chabrier \& Baraffe(1997)Chabrier \& Baraffe]{chabrierbaraffe1997}
         Chabrier, G., \& Baraffe, I\@.
         1997,
         \aap, 327, 1039

\bibitem[Chabrier et al\@.(2000)Chabrier et al\@.]{chabrieretal2000}
         Chabrier, G., Baraffe, I., Allard, F., \& Hauschildt, P\@.
         2000,
         \apj, 542, 464 

\bibitem[Dorman et al\@.(1998)Dorman et al\@.]{dormanetal1998}
         Dorman, B., Rood, R\@. T., \& O'Connell, R\@. W\@.
         1998,
         \apj, 419, 596

\bibitem[Edelmann(2003)Edelmann]{edelmann2003}
         Edelmann, H\@. 2003, 
         PhD thesis, 
         University Erlangen-Nuremberg,
         Germany

\bibitem[Fleig et al\@.(2008)Fleig et al\@.]{fleigetal2008} 
         Fleig, J., Rauch, T., Werner, K., \& Kruk, J\@. W\@.
         2008,
         \aap, 492, 565

\bibitem[Heber at al\@.(1984)Heber at al\@.]{heberetal1984}  
         Heber, U., Hunger, K., Jonas, G., \& Kudritzki, R\@. P\@.
         1984, 
         \aap, 130, 119

\bibitem[Hilditch et al\@.(1996)Hilditch et al\@.]{Hilditch96} 
         Hilditch, R\@. W., Harries, T\@. J., \& Hill, G\@. 
         1996, 
         \mnras, 279, 1380 

\bibitem[Hilditch et al\@.(2003)Hilditch et al\@.]{Hilditch03} 
         Hilditch, R\@. W., Kilkenny, D., Lynas-Gray, A\@. E., \& Hill, G\@. 
         2003, 
         \mnras, 344, 644 

\bibitem[Holweger(1979)Holweger]{holweger1979}        
         Holweger, H\@. 
         1979, 
         Les Elements et leurs Isotopes dans l'Universe,
         Universit\'e de Li\`ege, Institute de Astrophysique, p\@. 117

\bibitem[Hubeny et al\@.(1994)Hubeny et al\@.]{hea1994}
         Hubeny, I., Hummer, D\@. G., \& Lanz, T\@. 
         1994, 
         \aap, 282, 151

\bibitem[Hummer \& Mihalas(1988)Hummer \& Mihalas]{hm1988}
        Hummer, D\@. G., \& Mihalas, D\@. 
        1988, 
        \apj, 331, 794

\bibitem[Kilkenny(2011)Kilkenny]{kilkenny2011}      
         Kilkenny, D\@.
         2011,
         \mnras, 412, 487

\bibitem[Kudritzki(1976)Kudritzki]{kudritzki1976}      
         Kudritzki, R\@. P\@. 
         1976, 
         \aap, 52, 11

\bibitem[Kudritzki et al\@.(1982)Kudritzki et al\@.]{kudritzkietal1982}      
         Kudritzki, R\@. P., Simon, K\@. P., Lynas-Gray, A\@. E., Kilkenny, D., \& Hill, P\@. W\@. 
         1982, 
         \aap, 106, 254

\bibitem[Kurucz(1991)[Kurucz]{kurucz1991} 
         Kurucz, R\@. L\@. 
         1991, 
 	 in: Stellar Atmospheres: Beyond Classical Models, 
	 eds\@. L\@. Crivellari, I\@. Hubeny, D\@. G\@. Hummer, 
         NATO ASI Series C, Vol\@. 341,
         Kluwer, Dordrecht, p\@. 441 

\bibitem[Kurucz(2009)Kurucz]{kurucz2009}
         Kurucz, R\@. L\@.
         2009,
         in: Recent Directions in Astrophysical Quantitative
           Spectroscopy and Radiation Dynamics,
         eds\@. I\@. Hubeny, J\@. M\@. Stone, K\@. MacGregor, \& K\@. Werner, 
         AIP Conference Series Vol\@. 1171, p\@. 43

\bibitem[Lemke(1997)Lemke]{lemke1997}
         Lemke, M\@. 
         1997, 
         \aaps, 122, 285

\bibitem[M\"uller et al\@.(2010)M\"uller et al\@.]{muelleretal2010}	
	 M\"uller, S., Geier, S., \& Heber, U\@.
         2010,
         \apss, 329, 101

\bibitem[Napiwotzki \& Rauch(1994)Napiwotzki \& Rauch]{nr1994}
         Napiwotzki, R., \& Rauch, T\@. 
         1994,
        \aap, 285, 603

\bibitem[Rauch(2000)Rauch]{rauch2000} 
         Rauch, T\@. 
         2000, 
         \aap, 356, 665 

\bibitem[Rauch(2004)Rauch]{rauch2004}
         Rauch, T\@.
         2004,
         \apss, 291, 275

\bibitem[Rauch \& Deetjen(2003)Rauch \& Deetjen]{rd2003}
         Rauch, T., \& Deetjen, J\@. L\@. 
         2003,
         in: Stellar Atmosphere Modeling,
         eds\@. I\@. Hubeny, D\@. Mihalas, K\@. Werner,
         The ASP Conference Series, Vol\@. 288, p\@. 103

\bibitem[Rauch(1998)Rauch]{rauchetal1998} 
         Rauch, T\@., Dreizler, S., \& Wolff, B\@. 
         1998, 
         \aap, 338, 651 

\bibitem[Repolust et al\@.(2005)Repolust et al\@.]{repolust:2005}
         Repolust, T., Puls, J., Hanson, M\@. M., Kudritzki, R.-P., \& Mokiem, M\@. R\@.
         2005, 
         \aap, 440, 261
	
\bibitem[Tremblay \& Bergeron(2009)Tremblay \& Bergeron]{tremblaybergeron2009}
         Tremblay, P.-E., \& Bergeron, P\@.
         2009,
         \apj, 696, 1755

\bibitem[Vu{\v c}kovi{\'c} et al\@.(2008)Vu{\v c}kovi{\'c} et al\@.(]{vuckovicetal2008}
         Vu{\v c}kovi{\'c}, M., {\O}stensen, R., Bloemen, S., Decoster, I., \& Aerts, C\@.
         2008,
         in: Hot Subdwarf Stars and Related Objects,
         eds\@. U\@. Heber, C\@. S\@. Jeffery, R\@. Napiwotzki,
         The ASP Conference Series, Vol\@. 392, p\@. 199

\bibitem[Werner(1996)Werner]{werner1996}  
         Werner, K\@.
         1996,
         \apj, 457, L\,39

\bibitem[Werner et al\@.(2003)Werner et al\@.]{werneretal2003}  
         Werner, K., Dreizler, S., Deetjen, J\@. L., Nagel, T., Rauch, T., \& Schuh, S\@. L\@. 
         2003,
         in: Stellar Atmosphere Modeling,
         eds\@. I\@. Hubeny, D\@. Mihalas, K\@. Werner,
         The ASP Conference Series, Vol\@. 288, p\@. 31

\end{thebibliography}
\end{document}